\documentclass[prl,aps,showpacs,twocolumn,floatfix,superscriptaddress]{revtex4}
\usepackage{graphicx}
\usepackage{epsfig}

\newcommand{\beq}{\begin{equation}}
\newcommand{\eeq}{\end{equation}}
\newcommand{\beqs}{\begin{eqnarray}}
\newcommand{\eeqs}{\end{eqnarray}}

\def\hbar{\hspace{0pt}\raisebox{1pt}{$-$} \hspace{-7pt} h}

\begin{document}
\title{Neutrinos and $SU(3)$ Family
Gauge Symmetry}

\author{Thomas Appelquist \thanks{email: thomas.appelquist@yale.edu}}
\author{Yang Bai \thanks{email: yang.bai@yale.edu}}
\affiliation{Department of Physics, Sloane Laboratory, Yale
University, New Haven, CT 06520}
\author{Maurizio Piai \thanks{email:piai@u.washington.edu }}
\affiliation{Department of Physics, University of Washington,
Seattle, WA 98195}

\date{\today}

\begin{abstract}

We include the standard-model (SM) leptons in a recently proposed
framework for the generation of quark mass ratios and
Cabibbo-Kobayashi-Maskawa (CKM) mixing angles from an $SU(3)$ family
gauge interaction. The set of SM-singlet scalar fields describing
the spontaneous breaking is the same as employed for the quark
sector. The imposition at tree-level of the experimentally correct
Pontecorvo-Maki-Nakagawa-Sakata (PMNS) mixing matrix, in the form of
a tri-bi maximal structure, fixes several of the otherwise free
parameters and renders the model predictive. The normal hierarchy
among the neutrino masses emerges from this scheme.

\end{abstract}


\maketitle

\section{Introduction}

In a recent paper \cite{APY2}, we introduced an effective-field-theory (EFT)
framework for the computation of quark mass ratios and CKM mixing angles
based on an $SU(3)$ family gauge symmetry. Here we show that this framework
can accommodate the lepton masses and PMNS mixing angles through the
addition of only the three families of SM leptons, and that this fixes
several of the parameters of the model.

The EFT includes operators respecting the global family symmetries, coupling
SM-fermion bilinears and the Higgs-doublet to a set of SM-singlet scalars.
Spontaneous breaking of the family symmetries at a high scale then yields a
set of Goldstone-boson (GB) and pseudo-Goldstone-boson (PGB) degrees of
freedom, the symmetry being realized nonlinearly in the EFT below this scale
\cite{APY2}. The spontaneous breaking also leads to the Yukawa interactions
of the SM. Radiative corrections to these couplings arising from the $SU(3)$
family gauge interaction play a key role in generating realistic mass ratios
and mixing angles.

In addition, we include here a set of sub-leading operators which explicitly
break the global family symmetries. These operators affect small quantities
such as the up-quark mass and the CKM angle $\theta_{13}^q$. Most
importantly, they generate realistically small Majorana masses for the
neutrinos.

We first establish some notation and review the current experimental data.
We next describe our model, and discuss spontaneous breaking, masses, and
mixing angles at tree level (in the absence of family-gauge interactions).
We then include family-gauge radiative corrections, compare with
experimental data, and discuss our results.

~

~
\section{Fermion Masses and Mixing Angles}

\subsection{Notation}

Below the electroweak breaking scale, the quark and charged-lepton mass
operators are $-\, \bar{\psi}_L^{(i)}\,M^{(i)} \psi_R^{(i)}$, where
$i=u,d,e$. The $\psi_{L,R}^{(i)}$ are chiral fields for the quarks and
charged leptons and the $M^{(i)}$ are $3\,\times\,3$ matrices. Family
indices are understood. Similarly, with only the three left-handed neutrinos
present, the (Majorana) neutrino mass operator is $- \, (1/2) \,\nu_L^{T}\,
C\,M^{(\nu)}\,\nu_L$, where $C$ is the charge conjugation matrix.

All matrices are non-diagonal in flavor space and symmetric
in our specific case. One can
diagonalize them with appropriate transformations,
\beqs
\mbox{diag}\,M^{(i)}\,&=&\,L^{(i)\, \dagger}\,M^{(i)}\,L^{(i)\ast}\,,\\
\mbox{diag}\,M^{(\nu)}\,&=&\,L^{(\nu)\, T}\,M^{(\nu)}\,L^{(\nu)}\,, \eeqs
where  $L^{(i)}$ and $L^{(\nu)}$ are $3\times3$ matrices in flavor space.
The mixing matrices appearing in the charged-current weak interactions are
then given by \beqs
V_{CKM}\,&=&\,L^{(u)\,\dagger}\,L^{(d)}\label{ckm}\,,\\
V_{PMNS}\,&=&\,L^{(e)\,T}\, L^{(\nu)} \,, \label{pmns} \eeqs for quarks and
leptons, respectively. We use the standard definitions of the mixing
matrices, in which one writes the down-type quark (neutrino) flavor
eigenstates  $d$ ($\nu$) in terms of the mass eigenstates ${\hat d}$
(${\hat\nu}$)---in the basis in which up-type quarks (charged leptons) are
diagonal--- as \beqs
d & = & V_{CKM} \, {\hat d}\,,\\
\nu & = & V_{PMNS} \, {\hat \nu} \,. \eeqs

The standard parameterization of the Cabibbo-Kobayashi-Maskawa (CKM) matrix
in terms of three mixing angles $\theta_{12}^q$, $\theta_{13}^q$ and
$\theta_{23}^q$ and one  phase $\delta^q$ reads:
\begin{widetext}
\beqs
V_{CKM}\,=\,\left(\begin{array}{ccc}
c_{12}c_{13} & s_{12}c_{13} & s_{13}e^{-i\delta^{q}}\\
 - s_{12}c_{23} - c_{12}s_{23}s_{13}e^{i\delta^{q}} &
c_{12}c_{23} - s_{12}s_{23}s_{13}e^{i\delta^{q}} & s_{23}c_{13}\\
s_{12}s_{23} - c_{12}c_{23}s_{13}e^{i\delta^{q}} & -c_{12}s_{23} -
s_{12}c_{23}s_{13}e^{i\delta^{q}}
& c_{23}c_{13}\\
\end{array} \right)\,,
\eeqs
\end{widetext}
where $c_{ij}=\cos\theta_{ij}^q$ and $s_{ij}=\sin\theta_{ij}^q$.

An analogous expression in terms of $\theta_{ij}^l$ and $\delta^l$ is valid
for the Pontecorvo-Maki-Nakagawa-Sakata (PMNS) matrix $V_{PMNS}$, neglecting
the flavor-diagonal Majorana phases. The complete expression for the PMNS
matrix, including the Majorana phases $\phi_{1}$ and $\phi_{2}$
\cite{white}, reads $U_{PMNS}=V_{PMNS}\,K$ with
$K=$diag$\{1,e^{i\phi_{1}},e^{i(\phi_{2}+\delta^{l})}\}$. In our model,
Dirac and Majorana phases of order unity arise naturally, but all phases are
neglected here.

\subsection{Experimental data}

The quark masses in GeV units are $m_t(M_Z) = 176 \pm 5$, $m_b(M_Z) = 2.95
\pm 0.15$, $m_c(M_Z) = 0.65 \pm 0.12$, $m_s(M_Z) = 0.062 \pm 0.015$,
$m_u(M_Z) = 0.0017 \pm 0.0005$, $m_d(M_Z) = 0.0032 \pm 0.0009$. The CKM
mixing angles measured in tree-level processes, and as defined in Ref.
\cite{PDG}, are: $\sin \theta_{12}^q = 0.2243 \pm 0.0016$, $\sin
\theta_{23}^q = 0.0413 \pm 0.0015$, $\sin \theta_{13}^q = 0.0037 \pm
0.0005$.

The masses in GeV of the charged leptons are
 $m_{\tau}=1.78$,
$m_{\mu}=0.106$ and $m_e=0.511\times10^{-3}$, where we have neglected the
small errors. At present, combined fits of neutrino oscillation data
give~\cite{white}, at the $3\sigma$ level: \beqs \label{neutrinodata}
\Delta m^2_{12}& = & (7.1 - 8.9) \times 10^{-5}\, \mbox{eV}^2\,, \nonumber\\
|\Delta m^2_{23} | &= & (1.4 - 3.3) \times 10^{-3}\, \mbox{eV}^2\,, \nonumber\\
 \sin^2 2\,\theta_{12}^l & = &  0.70 - 0.94 \, ,  \nonumber \\
 \sin^2 \,\theta_{13}^l & < & 0.051\,,\nonumber\\
 \sin^2 2\, \theta_{23}^l  & = & 0.87 - 1.0 \, ,
\eeqs
where $\Delta m^2_{12}$ and $\Delta m^2_{23}$ are the solar and atmospheric
mass differences, respectively.

Two other observables, for which we have only upper bounds at present, are
the effective masses entering $\beta$-decay \beqs
m^2_{\beta}&\equiv&\sum_i |U_{ei}|^2m_i^2\,,\\
\label{betadecay} m_{\beta}&<&2.3\, \mbox{eV}\,, \eeqs and neutrinoless
double $\beta$-decay \beqs \label{nuless} m_{ee}&\equiv&\sum_i U_{ei}^2m_i\,,\\
\label{nulesslimit} |m_{ee}|&<&0.9\, \mbox{eV}\,, \eeqs where $m_i$ are the
eigenvalues of the neutrino mass matrix. In these expressions, $U_{ei}$ are
the elements of the complete $U_{PMNS}$. Hence $m_{ee}$ depends explicitly
on the phases, which are neglected here.

Each of these experimental quantities is measured at the electroweak
scale or below, while the parameters of our EFT are defined
naturally at the much higher family breaking scale. The parameters
must be evolved to the lower scales through SM interactions for a
precise comparison with experiment. These renormalization group (RG)
effects are expected to be small, and they depend on the choice of
the family breaking scale which we do not make here. We disregard
them and compare our expressions directly with the above quantities.

\section{The Model}

As in Ref.~\cite{APY2}, we introduce an $SU(3)$ family gauge symmetry,
taking it to be broken at some scale $F$, large enough to suppress
flavor-changing neutral currents. We employ an effective field theory (EFT)
including the SM interactions and the $SU(3)$ family gauge interaction to
describe physics below the cutoff $M_F \equiv 4 \pi F$. We take electroweak
breaking to be described by a Higgs-doublet field, requiring some additional
mechanism to stabilize the Higgs mass. We do not address this problem here.

With the family gauge coupling weak enough, the family gauge bosons
are part of the EFT, and their effects can be computed
perturbatively. The fermion fields of the EFT are those of the SM
quarks and leptons, together with a set of partners with the quantum
numbers of the up-type quarks. We assume that any additional new
physics, for example SM-singlet neutrinos or fields associated with
grand unification, appear only above the cutoff $M_F$.

The goal of Ref.~\cite{APY2} was to compute the quark mass ratios $m_d /
m_b$, $m_s / m_b$, $m_u /m_t$, $m_c / m_t$, and the CKM mixing angles
radiatively in the family gauge interaction. These quantities were arranged
to vanish in its absence by introducing two global symmetries, $SU(3)_1
\times SU(3)_2$, with the standard model fermions and their partners
transforming according to $SU(3)_1$ \cite{SU(3)refs}, and additional fields
of a "hidden" sector transforming according to $SU(3)_2$. The $SU(3)$ family
gauge interaction arose from gauging the diagonal subgroup of $SU(3)_1
\times SU(3)_2$. We also made use of an additional $Z_3$ symmetry to
classify the operators of the EFT.

This goal was only partly realized. While the quark mass ratios and
mixing angles were calculable perturbatively in the family gauge
interaction, and were expressed in terms of only two other small
parameters, the detailed results depended on a set of unknown ${\cal
O}(1)$ parameters. At a more detailed level, one of the CKM angles,
$\theta_{13}^q$,  as mentioned above, was well below the
experimentally allowed range.

Here, we extend this model to include the leptons by adding only the
$SU(2)_L$-doublet fields $\ell$ and the (charge conjugate) $SU(2)_L$-singlet
fields $e^c$, both transforming according to a $\bf{3}$ of $SU(3)_1$. The
fields of our model, together with their transformation properties under the
$SU(3)_1 \times SU(3)_2 \times Z_3$ symmetries and the SM symmetries, are
shown in Table~\ref{tab:fields}. Each fermion transforms as a $\bf{3}$ under
$SU(3)_1$ and therefore also as a $\bf{3}$ under the $SU(3)$ family gauge
symmetry. Thus the latter symmetry must be broken to generate fermion mass.
The fields $\chi$ and $\chi^c$, introduced in Ref.~\cite{APY2}, allow the
large mass hierarchies in the up-quark sector to be generated through a
seesaw mechanism.

As in Ref.~\cite{APY2}, in addition to the usual Higgs scalar $h$, two
SM-singlet scalars, $S$ and $\Sigma$, both $\bf{\bar{6}}$'s (symmetric
tensors) under $SU(3)_1$, couple to the fermions. The hidden sector is
described by one (SM-singlet) scalar multiplet, $H$, also a $\bf{\bar{6}}$
under $SU(3)_2$. In the EFT below the cutoff $M_F$, nonlinear constraints on
$S$, $\Sigma$, and $H$ insure that they describe only Goldstone-boson (GB)
and pseudo-Goldstone-boson (PGB) degrees of freedom. By contrast, the
electroweak Higgs sector is linearly realized.

         \begin{table}
\begin{tabular}{||c|c|c|c|c|c|c||}
          \hline
           & $SU(3)_1$ & $SU(3)_2$ & $Z_3$ & $SU(3)_c$ & $SU(2)_L$ & $U(1)_Y$
           \\ \hline \hline
         $q$  & 3 & 1 & $1''$ & 3 & 2 & $\frac{1}{6}$ \\ \hline
         $u^c$  & 3  &1  &$1'$  & $\bar{3}$ & 1 & $-\frac{2}{3}$ \\ \hline
         $d^c$  &3  & 1 & 1 & $\bar{3}$ & 1 & $\frac{1}{3}$ \\ \hline
         $\chi$  &  3 & 1 & $1'$ & 3 & 1 & $\frac{2}{3}$ \\ \hline
         $\chi^c$  & 3 & 1 & 1 & $\bar{3}$ & 1 & $-\frac{2}{3}$ \\ \hline
         $\ell$ & 3 & 1 & 1 & 1 & 2 & $-\frac{1}{2}$\\ \hline
         $e^c$ & 3 & 1 & $1''$ & 1 & 1 & 1 \\ \hline \hline
         $h$  & 1 &  1 & 1 & 1 & 2 & $-\frac{1}{2}$ \\  \hline
         $S$    & $\bar{6}$ & 1 & $1'$ & 1 & 1 & 0 \\  \hline
         $\Sigma$  & $\bar{6}$ & 1 & $1''$ & 1 &  1& 0 \\  \hline  \hline
         $H$  & 1 & $\bar{6}$ & 1 & 1 & 1 & 0 \\ \hline
          \hline
\end{tabular}
\caption{Field content and symmetries of the model. The $Z_{3}$
labels refer to the three cube roots of unity. All fermions are
left-handed chiral fields. The symbols $S$, $\Sigma$, and $H$ denote
SM-singlet scalar fields.} \label{tab:fields}
\end{table}

The $SU(3)$ family gauge interaction is universal with respect to all the
fermions. It is, so far, anomalous, requiring the existence of additional
heavy fermions to remove the anomalies. When integrated out, they generate
an appropriate Wess-Zumino-Witten (WZW) term in the EFT below $M_F$
\cite{WZW} which must be included in the analysis. It does not affect the
fermion mass ratios and mixing angles to leading order, and we do not
discuss it further.

\section{Tree Level}

We first discuss the structure of the model at tree level, i.e. in the
absence of the $SU(3)$ family gauge interaction. The tree-level Yukawa
Lagrangian for the fermion masses and mixing angles consists, first of all,
of terms that respect the global symmetries of the model. In Ref.
\cite{APY2}, only such terms were considered. Here, in addition to
incorporating the leptons, we also include smaller, symmetry breaking terms
in the tree-level Lagrangian. These affect quantities that were estimated to
be very small in  Ref.~\cite{APY2}, for example the CKM angle
$\theta_{13}^q$ which was well below the experimental range. They also play
a central role in the generation of neutrino masses and mixing angles.

The symmetry-preserving terms are given by \beqs \label{Lag-tree}
-{\cal L}_Y &=&
y_d\frac{qhSd^c}{F}+y_1\frac{q\tilde{h}S\chi^c}{F}+y_2\chi
Su^c+y_3\chi\Sigma \chi^c\nonumber\\
&&+ y_e\frac{\ell hSe^c}{F}+{\rm h.c.}
         \eeqs
They are invariant under $SU(3)_1 \times SU(3)_2$ and $Z_3$, as well as
$U(1)_q \times U(1)_l$ describing quark and lepton number conservation, and
the SM gauge symmetries. The scalars $S$ and $\Sigma$ are neutral under
$U(1)_q \times U(1)_l$. (We assume that the $U(1)_S$ and $U(1)_{\Sigma}$
symmetries associated with these complex fields are broken explicitly by the
underlying dynamics \cite{APY2}.) Each of the $y_i$ couplings is a
dimensionless parameter determined by physics above $M_F$. Each except for
$y_1$ is small compared to the family gauge coupling $g$, which will be
${\cal O}(1)$, that is, $\alpha/\pi \equiv g^2/4\pi^2 = {\cal O}(1/40)$.
This will allow using the $y_i$ couplings at only first order, with quantum
corrections arising from the family gauge interactions alone.

A comment is in order to explain why Eq.~(\ref{Lag-tree}) contains
only five interactions, all linear in $S$ and $\Sigma$. After all,
there are many invariant operators bilinear in the fermion fields,
but with higher powers of $S$ and $\Sigma$. With the scale of these
operators taken to be the same as the VEV's of $S$ and $\Sigma$ as
we do here, they are not surpressed by power counting. In the limit
$y_i\rightarrow 0$, however, this Lagrangian preserves an
additional, global $U(1)$ for each of the $y_i$. It is hence
technically natural to assume that each of the $y_i$ is a small
parameter. At the loop level, combinations of the couplings in
Eq.~(\ref{Lag-tree}) together with scalar self-interactions generate
all other possible operators compatible with the exact symmetries,
containing higher powers of $S$ and $\Sigma$, and still bilinear in
the fermion fields. But the coefficients of operators generated in
this way are suppressed by products of the $y_i$, and are hence very
small. Their effects are comfortably below those induced by the
gauge interaction discussed here. Eq.~(\ref{Lag-tree}) represents
the first few terms of an expansion, truncated at the leading order
in the global $U(1)$ symmetry-breaking, small parameters $y_i$.

A word about anomalies of global symmetries is in order. Fist of
all, we disregard the familiar $B+\ell$ anomaly induced by the
electroweak interactions since its effects are very small. There is
also a global $U(1)_l$ anomaly induced by the $SU(3)$ family gauge
interaction. While this could be canceled by physics above $M_F$
(for example, a set of SM-singlet neutrinos), the $U(1)_l$ symmetry
is broken in the EFT itself.

We turn next to the additional Yukawa operators that explicitly
break the global symmetries $SU(3)_1 \times SU(3)_2$ and/or $Z_3$ by
small amounts. These terms are taken to preserve $U(1)_q$. (We
assume that the $U(1)_H$ symmetry associated with the complex field
$H$ is broken explicitly by the underlying dynamics.)

 We include the following small symmetry-breaking operators: \beqs
\label{BreakingLag-tree} -{\cal L}_Y^{\prime} =
y_u^{\prime}\frac{q\tilde{h}\Sigma u^c}{F} + y^{\prime}_e\frac{\ell
h\Sigma
e^c}{F}+\frac{y_{\nu}^{\prime}}{2}\frac{\ell\tilde{h}H\tilde{h}\ell}{F^2}+{\rm
h.c.}.
         \eeqs
The first contributes to the up-quark masses and (importantly) to
$\theta_{13}^q$, and the second contributes to the charged-lepton
masses. The last term is the source of the (small) Majorana mass
matrix for the neutrinos. The first two terms break $Z_3$, while the
third breaks $SU(3)_1 \times SU(3)_2$ to its diagonal subgroup. The
final breaking is also triggered by the $SU(3)$ family gauge
interaction. Each of the dimensionless coefficients in $ {\cal
L}_Y^{\prime}$ will be of ${\cal O}(10^{-4})$ or smaller, well below
those in ${\cal L}_Y$.

Additional symmetry-breaking operators can also be included. An
obvious example is $qh\Sigma d^c/F$, similar in structure to the
first two terms above, and contributing to the down-quark masses.
The coefficients of these operators can consistently be taken to be
very small, since, when they are generated from quantum loops based
on the interactions of Eqs.~(\ref{Lag-tree}) and
(\ref{BreakingLag-tree}), they arise with very small coefficients.
These operators then produce very small physical effects.

\subsection{Spontaneous Breaking}

Here, as in Ref.~\cite{APY2}, we assume that the global symmetries
$SU(3)_1 \times SU(3)_2$ and $Z_3$ are broken spontaneously at the
scale $F$ by vacuum expectation values (VEV's) of the scalar fields
$S$, $\Sigma$ and $H$. The VEV of $H$ also breaks $U(1)_l$. The
VEV's are taken to be \beqs
       \label{tree-vev}
\langle S\rangle & = & F\left(
              \begin{array}{ccc}
               0 & 0 & 0 \\
                0 & 0 & 0 \\
                0 & 0 & s \\
              \end{array}
            \right)\,,\\
\langle\Sigma\rangle & = & F\left(
              \begin{array}{ccc}
                0 & 0 & 0 \\
                0 & \sigma & 0 \\
                0 & 0 & 0 \\
              \end{array}
            \right)\,,\\
         \label{ab}
\langle H\rangle & = & F\left(
              \begin{array}{ccc}
              b_1^2 & b_2 & b_3 \\
               b_2 & a_1 & a_2 \\
               b_3 & a_2 & 1 \\
              \end{array}
            \right)\,,
            \eeqs
where $s$, $\sigma$ and the $a_i$ will be ${\cal O}(1)$, while the $b_i$
will be of order the Cabibbo angle $\theta_{12}^q$. The overall scale of
these dimensionless parameters can be absorbed into the definition of $F$,
so we have arbitrarily set one element of $\langle H\rangle / F$ to unity.
We neglect all phases, taking each of the dimensionless parameters to be
positive. It was argued in Ref.~\cite{APY2} that the above pattern in the
visible sector (the $S$ and $\Sigma$ fields), and the hierarchical structure
in the "hidden" sector (the $H$ field), emerge naturally from a class of
potentials.

In the visible sector, since both $\langle S\rangle$ and
$\langle\Sigma\rangle$ are diagonal, two $Z_2$ subgroups of $SU(3)_1$ are
preserved. The quark mixing angles vanish in the absence of the family gauge
interaction. The breaking pattern in the hidden sector, being described by a
single sextet $H$ field, automatically preserves two $Z_2$ subgroups of
$SU(3)_2$. (This is most evident in a frame in which $\langle H\rangle$ is
made diagonal.) The alignment of the visible and hidden sectors will be
determined dominantly by the family gauge interaction which links them. We
assume here that these interactions misalign the two sectors such that no
$Z_2$ symmetry remains when the sectors are gauge coupled. In the limit $b_i
\rightarrow 0$, an exact $Z_2$ remains in tact.

The spontaneous breaking of $SU(3)_1 \times SU(3)_2$ leads to a set
of $8$ PGB's along with the massive gauge bosons. The PGB's acquire
mass from explicit symmetry breaking, dominantly due to the gauge
interaction, of ${\cal O}(g^{2}F/4\pi)$, and are thus part of the
EFT. But they couple weakly to the fermions, and are neglected here
\cite{APY2}.

\subsection{Mass Matrices}

After electroweak symmetry breaking, the mass matrices for the
fermions are generated by ${\cal L}_Y$ Eq.~(\ref{Lag-tree}) and
${\cal L}_Y^{\prime}$ Eq.~(\ref{BreakingLag-tree}) with the scalar
fields replaced by their VEVs. For down-type quarks,
  \beqs \label{down}
         M^{(d)}= y_d v \frac{\langle S \rangle}{F} = y_d\,v\left(
              \begin{array}{ccc}
               0 & 0 & 0 \\
                0 & 0 & 0 \\
                0 & 0 & s \\
              \end{array}
            \right),
         \eeqs
         where $v \simeq 250 GeV$ is the VEV of Higgs doublet $h$. At this
          level, only the
         $b$ quark develops a mass, of the right order for $y_d \simeq
         10^{-2}$.

       The up-type quark mass matrix is $6\times 6$:
         \beqs
         \label{seesaw}
(u\;\chi) \tilde{M}^{(u)} \left(
          \begin{array}{c} u^c \\ \chi^c \\ \end{array} \right)
         =(u\;\chi) \left(
                    \begin{array}{cc}
                       y_u^{\prime}v\frac{\langle \Sigma\rangle}{F} & y_1\,v\frac{\langle S\rangle}{F} \\
                      y_2\langle S\rangle &y_3\langle\Sigma\rangle \\
                    \end{array}
                  \right)
         \left(  \begin{array}{c}u^c \\\chi^c \\
          \end{array}\right).
         \eeqs
The squares of the eigenvalues of this (non-symmetric) matrix can be read
off from the diagonal matrix $\tilde{M}^{(u)} \tilde{M}^{(u)\dagger}$. There
are four non-vanishing eigenvalues. Two are small ($y_1^{2}v^2$ and
$y_u^{2}v^2$), and two are large ($y_2^{2}F^2$ and $y_3^{2}F^2$) providing
that $v/F \ll y_2, y_3$. When the family gauge interactions are included,
another large eigenvalue is generated, and a seesaw mechanism leads to
masses and mixing angles for the up-type quarks.

For charged leptons, the mass matrix is
\beqs
         M^{(e)}&=& y_e v \frac{\langle S \rangle}{F} + y^{\prime}_e v \frac{\langle \Sigma \rangle}{F} \nonumber\\
         &=&y_e\,v\left(
              \begin{array}{ccc}
               0 & 0 & 0 \\
                0 & \frac{y_e^{\prime}}{y_e} \sigma & 0 \\
                0 & 0 & s \\
              \end{array}
            \right),
         \eeqs
from which one can read off the (tree-level) masses $m_{\tau}=y_e s v$ and
$m_{\mu}=y_e^{\prime} \sigma v$. At this level, the electron is massless and
the lepton mixing matrix $L^{(e)}$ is the identity matrix.

For neutrinos, the Majorana mass matrix is proportional to the VEV of $H$:
\beqs \label{majoranamass}
         M^{(\nu)}&=&  \frac{y_{\nu}^{\prime} v^2}{F} \frac{\langle H \rangle}{F} \nonumber\\
         &=& \frac{y_{\nu}^{\prime}v^2}{F}\left(
              \begin{array}{ccc}
              b_1^2 & b_2 & b_3 \\
               b_2 & a_1 & a_2 \\
               b_3 & a_2 & 1 \\
              \end{array}            \right).
         \eeqs
Thus the scalar field $H$ plays a more central role here than in Ref.
\cite{APY2} where it entered only through its family-gauge coupling.

\subsection{Neutrinos}
We next observe that the pattern of neutrino masses and mixing
angles can be accommodated at tree level, placing a further
restriction on the above parameters. We first recall that there will
be a hierarchical structure to $\langle H\rangle$: $b_i = {\cal
O}(b) < a_j = {\cal O}(1)$. A simple approach is to impose further
on $\langle H\rangle$ a form that leads to tri-bi-maximal mixing
\cite{tri-bi}. This so far provides a good fit to the experimentally
measured angles. At tree level, since the charged lepton matrix
$L^{(e)}$ remains the identity, the tri-bi form corresponds to
 \beq \label{tri-bi}
L^{(\nu)}= \left(\begin{array}{ccc}
 \sqrt{\frac{2}{3}} & \frac{1}{\sqrt{3}} & 0 \\
    -\frac{1}{\sqrt{6}}   & \frac{1}{{\sqrt{3}}} & \frac{1}{{\sqrt{2}}} \\
    \frac{1}{\sqrt{6}} & - \frac{1}{\sqrt{3}} & \frac{1}{\sqrt{2}} \\
      \end{array} \right), \eeq
leading to the PMNS mixing angles $\theta_{23}^l = \pi/4$, $\theta_{12}^l =
\sin^{-1}\sqrt{1/3}$, and $\theta_{13}^l = 0$.

The tri-bi form emerges by imposing on the parameters of $\langle H \rangle$
the three conditions \beq \label{tri-birest} b_3 = -b_2~,~ a_1 = 1~,~ a_2 =
1 - b_2 - b_1^2. \eeq These relations do not correspond to a symmetry limit
of our model, and we have not obtained them by a vacuum alignment analysis.
We impose them simply to accommodate the current experimental data on the
PMNS angles, reducing the number of parameters used to describe the quarks,
charged leptons, and neutrino-mass hierarchy. With these conditions,
$\langle H \rangle$ takes the form \beqs \frac{\langle H \rangle}{F}
&=&\left(\begin{array}{ccc}
     b_1^2 & b_2 & -b_2 \\
     b_2 & 1 & 1-b_2-b_1^2 \\
     -b_2 & 1-b_2-b_1^2 & 1 \\
   \end{array}
 \right) \nonumber \\
 &=&L^{(\nu)}
\left(\begin{array}{ccc}
 b_1^2-b_2 & 0 & 0 \\
    0   & b_1^2+2b_2& 0 \\
    0 & 0 & 2-b_1^2-b_2 \\
      \end{array}
 \right)
 L^{(\nu)T}\nonumber \\
 \eeqs

With $b_i \ll 1$, we then have \beqs \label{m1}
 m_1 \simeq -b_2\frac{y_{\nu}^{\prime} v^2}{F} \\ \label{m2}
 m_2 \simeq 2b_2\frac{y_{\nu}^{\prime} v^2}{F}  \\
 m_3 \simeq 2\frac{y_{\nu}^{\prime} v^2}{F}  \,,
  \eeqs
and therefore \beqs
 \Delta m_{23}^2 & \simeq & 4\frac{y_{\nu}^{\prime 2} v^4}{F^2} \label{deltam23}\\
\Delta m_{12}^2 & \simeq & 3b_2^2\frac{y_{\nu}^{\prime2 } v^4}{F^2}.
  \eeqs
Thus, the normal neutrino hierarchy emerges from the imposition of the
tri-bi form on $\langle H \rangle$.

There is a simple reason for this. Without the tri-bi conditions,
Eq.~(\ref{tri-birest}), $\langle H \rangle$ has one small (${\cal
O}(b)$) eigenvalue and two large eigenvalues, with the size of $b$
being set by the Cabibbo angle $\theta_{12}^q$. The tri-bi
conditions (together with the hierarchy $b_i < a_j$) restrict the
$2-3$ sub-matrix to have a small determinant: $a_1 - a_2^2 = {\cal
O}(b)$. There are then two small eigenvalues and one large
eigenvalue, a consequence of only this restriction, not requiring
the full imposition of the tri-bi conditions. The hierarchy between
solar and atmospheric mass-squared differences follows
automatically.

The restriction $a_1 - a_2^2 = {\cal O}(b)$, by itself, gives also
$\theta_{13}^l = {\cal O}(b)$, that is, $\sin^2\theta_{13}^l = {\cal
O}(b^2)$ at tree level. The full set of tri-bi conditions,
Eq.~(\ref{tri-birest}), leads to the vanishing of $\theta_{13}^l$,
but this tree-level result will be lifted by the radiative
corrections, leading to an ${\cal O}(b^2)$ estimate. We note finally
that the single restriction $a_1 - a_2^2 = {\cal O}(b)$ also leads
to an approximate $U(2)$ invariance of  $\langle H \rangle$.

\section{Radiative Corrections to the Mass Matrices}

The tree-level theory with the conditions Eq.~(\ref{tri-birest}) can
accommodate a realistic neutrino mass matrix while yielding vanishing CKM
angles and vanishing masses for the first-family quarks. We next discuss the
effects of the $SU(3)$ family gauge interaction. Perturbation theory will be
valid since $g^{2}/4\pi^2 \ll 1$. The loop corrections to the mass matrices
may be viewed as corrections to $\langle S \rangle$, $\langle \Sigma
\rangle$ and $\langle H \rangle$.

At one loop, we find \beq \label{radiative} \delta \langle S
\rangle_{ij}\,=\,-\frac{\alpha}{\pi}sF
\log(\frac{M^2_c}{M_F^2})(t_a)^3_i(t_b)^3_j O_{ac}O_{bc},
      \eeq
where $i,j = 1,2,3$ are the family indices and $a,b,c = 1,\cdots,8$
label the 8 gauge bosons. $M_F \equiv 4\pi F$ is the cutoff scale
and the $M_c^2$ are the mass eigenvalues of the family gauge bosons.
The matrix $O$ is the orthogonal transformation diagonalizing the
gauge boson mass matrix. The small parameters $b_1$ and $b_2$ enter
this matrix. The $M_F$ dependence survives in only the $33$ element.
Similar expressions obtain for $\delta \langle \Sigma \rangle_{ij}$
and $\delta \langle H \rangle_{ij}$. In $\delta \langle \Sigma
\rangle_{ij}$, the $M_F$ dependence enters only the $22$ element. In
$\delta \langle H \rangle_{ij}$, the $M_F$-dependent term is
proportional to $\langle H \rangle_{ij}$. Thus, at one loop level in
the gauge interaction, the cutoff ($M_F$) dependence is universal,
and can be absorbed into a renormalization of the coupling constants
in ${\cal L}_Y$ and ${\cal L}_Y^{\prime}$.

To derive these expressions, it is convenient to work in a renormalizable
gauge. Eq.~(\ref{radiative}) is derived in Feynman gauge. There are also
corrections due to wave function renormalization (kinetic energy mixing)
arising from the family gauge interaction, as well as contributions from
emission and re-absorption of the GB degrees of freedom. They lead to
corrections of the same general form with no new parameters, and we do not
exhibit them explicitly.

The corrected form of the $\langle S \rangle$ matrix thus includes ${\cal
O}(\alpha/\pi)$ entries replacing the $0$'s in Eq.~(\ref{tree-vev}). The
presence of the small parameters $ b_1 $ and $ b _2$ in the first row and
column of $\langle H \rangle$ leads through its contribution to the
gauge-boson masses to a similar presence in the corrected $\langle S
\rangle$. Its general form is
         \beqs \label{Sprime}
\langle S\rangle'=\langle S\rangle + \delta \langle S \rangle = F\left(
              \begin{array}{ccc}
               {\cal
O}(\frac{\alpha}{\pi} b ^2) & {\cal O}(\frac{\alpha}{\pi} b ) &{\cal
O}(\frac{\alpha}{\pi} b ) \\
               {\cal
O}(\frac{\alpha}{\pi} b ) &{\cal O}(\frac{\alpha}{\pi}) & {\cal
O}(\frac{\alpha}{\pi}) \\
               {\cal
O}(\frac{\alpha}{\pi} b ) &{\cal
O}(\frac{\alpha}{\pi}) & s \\
              \end{array}
            \right).\eeqs
Similarly, the general form of the corrected $\langle \Sigma \rangle$ matrix
is \beqs \langle \Sigma \rangle'=\langle \Sigma \rangle + \delta \langle
\Sigma \rangle = F\left(
              \begin{array}{ccc}
             {\cal
O}(\frac{\alpha}{\pi} b ^2)    & {\cal O}(\frac{\alpha}{\pi} b ) & {\cal
O}(\frac{\alpha}{\pi} b ) \\
                {\cal
O}(\frac{\alpha}{\pi} b ) & \sigma & {\cal
O}(\frac{\alpha}{\pi}) \\
               {\cal
O}(\frac{\alpha}{\pi} b ) & {\cal O}(\frac{\alpha}{\pi}) & {\cal
O}(\frac{\alpha}{\pi}) \\
              \end{array}
            \right).
         \eeqs
Here again, each entry in the first row and column carries a suppression
factor of ${\cal O}( b )$.

The precise form of $\langle S \rangle'$, used in place of its tree-level
counterpart in Eq.~(\ref{down}), then leads to the corrected down-type quark
mass matrix $M^{(d) \prime}$. Similarly, the precise $\langle S \rangle'$
and $\langle \Sigma \rangle'$, used in Eq.~(\ref{seesaw}), lead to a
corrected $6 \times 6$ up-type matrix. After integrating out the $3$ heavy
up-type fermions (implementing the up-sector seesaw), we obtain the $3
\times 3$ matrix $M^{(u) \prime}$ for the $u$, $c$, and $t$ quarks. The
masses of the heavy up-type fermions are of order $y_2 F$, and $y_3 F$, and
$(\alpha/\pi)y_{3}b F$. These are below the cutoff $M_F$ but well above the
electroweak scale.

The matrices $\langle S \rangle'$ and $\langle \Sigma \rangle'$ also enter
the charged-lepton mass matrix, which takes the general form \beqs
  M^{(e)\prime}&=& y_e v \frac{\langle S \rangle^{\prime}}{F} + y^{\prime}_e v \frac{\langle \Sigma \rangle^{\prime}}{F} \nonumber\\
         &=&y_e\,v\,s\left(
              \begin{array}{ccc}
                 {\cal
O}(\frac{\alpha}{\pi} b ^2) & {\cal O}(\frac{\alpha}{\pi} b ) &{\cal
O}(\frac{\alpha}{\pi} b ) \\
               {\cal
O}(\frac{\alpha}{\pi} b ) &{\cal O}(\frac{\alpha}{\pi})+
\frac{y_e^{\prime} \sigma}{y_e s}  & {\cal
O}(\frac{\alpha}{\pi}) \\
               {\cal
O}(\frac{\alpha}{\pi} b ) &{\cal
O}(\frac{\alpha}{\pi}) & 1
                           \end{array}
            \right).
         \eeqs

We note that the matrix $L^{(e)}$ which diagonalizes $M^{(e)\prime}$
and enters $V_{PMNS}$, contributes terms of ${\cal O}(b)$, not
suppressed by ${\cal O}(\alpha/\pi)$, to the mixing in the $1-2$
sub-sector. Its general form, neglecting terms of ${\cal
O}(\alpha/\pi)$, is \beqs \label{Leapprox} L^{(e)}&\simeq& \left(
              \begin{array}{ccc}
               1 &  {\cal O}(b)  & 0 \\
                - {\cal O}(b)  & 1& 0 \\
                0 & 0 & 1 \\
              \end{array}
            \right)\,,
\eeqs This same feature appears in the CKM matrix, that is,
$\theta_{12}^q = {\cal O}(b)$. In the present case, it relies on the
fact that the quantity $y_e^{\prime} \sigma / y_e s$ turns out to be
no larger than ${\cal O}(\alpha/\pi)$.

The approximate form Eq.~(\ref{Leapprox}) leads to an approximation
for the PMNS matrix $V_{PMNS} \equiv L^{(e)T} \ L^{(\nu)}$. The
neutrino matrix $M^{(\nu)}$ is given by Eq.~(\ref{majoranamass})
together with its one-loop radiative corrections. As we have noted,
these can be viewed as corrections $\delta \langle H \rangle_{ij}$
to $\langle H \rangle_{ij}$, and are all of ${\cal O}(\alpha/\pi)$.
If they are neglected, then with the tri-bi conditions,
Eq.~(\ref{tri-birest}), $L^{(\nu)}$ continues to have the form Eq.
(\ref{tri-bi}), and $V_{PMNS}$ has the same form together with
corrections of ${\cal O}(b)$. This leads to small shifts to the
tri-bi values $\sin^2{2\theta_{12}^l} = 8/9$ and
$\sin^2{2\theta_{23}^l} = 1$, and to \beqs \label{13}
\sin^2\theta_{13}^l = {\cal O}(b^2)\,. \eeqs This result emerged
already at tree level if only the normal neutrino hierarchy was
imposed on $\langle H \rangle$.

\section{Phenomenology}

There are four adjustable Yukawa couplings, determined by physics
above the cutoff $M_F$, that set the scale for the up-type quarks,
the down-type quarks, the charged leptons, and the neutrinos. The
first three are $y_1 = {\cal O}(1)$, $y_d = {\cal O}(10^{-2})$, and
$y_e = {\cal O}(10^{-2})$, which enter the symmetry-preserving
Lagrangian Eq.~(\ref{Lag-tree}). The coupling $y_{\nu}^{\prime}$
enters the symmetry-breaking Lagrangian Eq.
(\ref{BreakingLag-tree}). The expression for $\Delta m_{23}^2$, Eq.
(\ref{deltam23}) leads to $y_{\nu}^{\prime}= {\cal
O}(F/10^{15}GeV)$. Thus $y_{\nu}^{\prime} \ll 1 $ providing that
$F$, the family-breaking scale, is well below the grand unified
theory scale. We won't need to commit to a particular value for $F$,
except, as noted above, that it must be large enough to suppress
flavor-changing neutral currents.

We focus here on the remaining parameters of the model
(Table~\ref{tab:input}), employing them to reproduce the fermion
mass ratios and mixing angles. There are 2 ${\cal O}(1)$ parameters
$s$ and $\sigma$, entering the VEV's $\langle S \rangle$ and
$\langle \Sigma \rangle$. There are the 2 parameters $b_1$ and
$b_2$, entering $\langle H \rangle$, which will be comparable and
small (${\cal O}(b)$), describing the small spontaneous breaking of
a $Z_2$ symmetry. The size of $b$ is set by the Cabibbo angle
$\theta_{12}^q$. The gauge coupling $\alpha/\pi$ is even smaller,
breaking $SU(3)_1 \times SU(3)_2$ to its diagonal subgroup. The two
Yukawa couplings $y_2$ and $y_3$ in ${\cal L}_Y$ enter only as the
ratio $z \equiv y_2/y_3$ in the seesaw expression for the up-type
quark masses. We will find $z$ to be a small parameter. Finally,
there are the two small ratios $z_u = y_u^{\prime}/y_1$ and $z_e =
y_e^{\prime}/y_e$ arising from the symmetry-breaking Lagrangian
${\cal L}_Y^{\prime}$ Eq.~(\ref{BreakingLag-tree}).

The restriction to just these parameters is due to the imposition of
the conditions Eq.~(\ref{tri-birest}) leading at tree-level to the
form of the PMNS matrix. There are then $10$ experimental quantities
(Table~\ref{tab:output}) to be accommodated after the inclusion of
the radiative corrections. They are the up-type mass ratios
$m_u/m_t$ and $m_c/m_t$, the down-type ratios $m_d/m_b$ and
$m_s/m_b$, the CKM angles $\theta_{12}^q$, $\theta_{23}^q$, and
$\theta_{13}^q$, the charged-lepton mass ratios $m_e/m_{\tau}$ and
$m_{\mu}/m_{\tau}$, and the neutrino mass ratio $\Delta m_{12}^2 /
\Delta m_{23}^2$.

To check that the model effectively reproduces the experimental
data, we perform a numerical study by sampling the parameter space.
An example of a set of parameters that reproduces the data well is
reported in Table~\ref{tab:input}. For the study it is sufficient to
use the tree-level neutrino mass matrix. Radiative corrections are
crucial, however, for the quarks and charged leptons. The radiative
expressions in terms of the $8$ parameters, are in general somewhat
complicated. For orientation, we exhibit the approximate algebraic
form of these expressions in the quark sector for the parameter
values that emerge from the numerical study (Table~\ref{tab:input}).
For values in this range, it can be shown that \beqs  \theta_{12}^q
= {\cal O}(b)~,~ \theta_{23}^q = {\cal O}(\frac{\alpha}{\pi}) ~,~
\theta_{13}^q = {\cal O}(\frac{bz_u}{z})\, \, , \eeqs \beqs
\frac{m_c}{m_t} = {\cal O}( \frac{z}{b}\frac{\alpha}{\pi})~,~
\frac{m_u}{m_t} = {\cal O}(b \sqrt{z z_u}\frac{\alpha}{\pi})\, ,
\eeqs \beqs \frac{m_s}{m_b} = {\cal O}(\frac{\alpha}{\pi})~,~
\frac{m_d}{m_b} = {\cal O}(\frac{\alpha}{\pi}b^2) \, . \eeqs  The
expressions for $\theta_{13}^q$ and $m_u/m_t$ depend directly on the
small parameter $z_u$ which measures the strength of the
symmetry-violating operator $q \tilde{h} \Sigma u^c/F$ in ${\cal
L}_Y^{\prime}$. It was neglected in Ref.~\cite{APY2}.

Two similar algebraic formulas obtain for the charged-lepton mass ratios
$m_e/m_{\tau}$ and $m_{\mu}/m_{\tau}$. Their dependence on the parameter
$z_e = y_e^{\prime}/y_e$ is essential to accommodate the fact that they
differ from the down-type ratios $m_d/m_b$ and $m_s/m_b$

The radiative corrections $\delta \langle S \rangle_{ij}$ and
$\delta \langle \Sigma \rangle_{ij}$ depend on the cutoff $M_F$, but
this dependence enters only the $33$ and $22$ elements respectively.
As already noted, it can be interpreted as a renormalization of the
Yukawa couplings in Eqs. (\ref{Lag-tree}) and
(\ref{BreakingLag-tree}). It does not enter the expressions for the
mass ratios and mixing angles. The one-loop calculations, with
$\alpha/\pi \simeq 0.05$, are no more accurate than about $5\%$. In
addition, we are neglecting RG running effects from the symmetry
breaking scale $F$ to the electroweak scale, which can affect the
comparison with experiment at the same level. Some further
uncertainties affecting the first family arise from having neglected
phases.

\begin{table}[!h]
\begin{ruledtabular}
\begin{tabular}{|c|c|c|c|}
                  $b_1$ & $b_2$ &  $s$ & $\sigma$
                 \\ \hline
                0.16 & 0.17 & 0.62   & 0.55
                \\ \hline \hline
                 $\alpha/\pi$ & $z$ & $z_e$ & $z_u$  \\ \hline
                 0.053 & 0.014 & 0.055 &  0.00028  \\
             \end{tabular}
\end{ruledtabular}
\caption{Input values of the 8 parameters, which are taken to be
real and positive.}
\label{tab:input}
\end{table}

\begin{table}[ht]
\begin{ruledtabular}
\begin{tabular}{|c||c|c|}
& Exp. & Model \\ \hline
  \hline
  $\frac{m_u}{m_t}$ & $9.8\pm 5.4\times 10^{-6}$ & $5.7\times 10^{-5}$ \\ \hline
  $\frac{m_c}{m_t}$ & $0.0037\pm 0.0008$ & 0.0030 \\ \hline
  $\frac{m_d}{m_b}$ & $0.0011\pm 0.0006$ & 0.0021 \\ \hline
  $\frac{m_s}{m_b}$ & $0.021\pm 0.006$ &  0.014\\ \hline
  $\sin\theta_{12}^q$ & $0.2243 \pm 0.0016$ & 0.22 \\ \hline
  $\sin\theta_{23}^q$ & $0.0413\pm 0.0015$ & 0.042 \\ \hline
  $\sin\theta_{13}^q$ & $0.0037\pm 0.0005$ & 0.0037 \\ \hline
  $\frac{m_e}{m_\tau}$ & $2.88\times 10^{-4}$ & $2.9 \times 10^{-4}$ \\ \hline
  $\frac{m_\mu}{m_\tau}$ & $0.0595$ & $0.060$\\ \hline
$\frac{\Delta m^2_{12}}{\Delta m^2_ {23}}$ & $0.034\pm 0.013$ &  0.034 \\
\end{tabular}
\end{ruledtabular}
\caption{Comparison to the 10 experimental quantities. The accuracy
of the values from the model are limited by uncertainties associated
with perturbation theory, RG evolution to the electroweak scale and
the neglect of phases. The small experimental errors for
$m_e/m_\tau$ and $m_\mu/m_\tau$ are neglected.}
 \label{tab:output}
\end{table}

\section{Discussion}

Most of the 10 small experimental quantities of
Table~\ref{tab:output} are reproduced well by the 3 small parameters
$b_1$, $\alpha/\pi$, and $z \equiv y_2/y_3$ , together with the 3
${\cal O}(1)$ parameters $s$, $\sigma$ and $b_2/b_1$. The smallest
angle $\theta_{13}^q$ and the smallest mass ratio $m_u/m_t$ require
the inclusion of the small parameter $y_u^{\prime}$ ($z_u =
y_u^{\prime}/y_1$) from the symmetry-breaking Lagrangian ${\cal
L}_Y^{\prime}$, Eq.~(\ref{BreakingLag-tree}). The CKM angle
$\theta_{13}^q$, which was much too small in Ref.~\cite{APY2} where
the small symmetry breaking operators were neglected, is now in the
experimentally allowed range. A more accurate treatment of
$\theta_{13}^q$ would call for the inclusion of the CP-violating
phase $\delta^q$ in $V_{CKM}$. The ratio $m_u/m_t$ turns out to be
somewhat large, but this is the smallest experimental quantity
considered, and therefore most sensitive to additional small
corrections. The charged-lepton mass ratios are both moderately
sensitive to $y_e^{\prime}$ ($z_e = y_e^{\prime}/y_e$), also
entering ${\cal L}_Y^{\prime}$.

We note that the numerical values of $z_e$ and $z_u$ correspond to
$y_e^{\prime} = {\cal O}(10^{-4})$ and $y_u^{\prime} = {\cal
O}(10^{-4})$. Since $y_{\nu}^{\prime}= {\cal O}(F/10^{15}GeV)$, this
coupling constant, too, will be at least this small providing only
that $F \leq {\cal O}(10^{11}GeV)$. Then, for a range of values of
$y_3$ in the symmetric Lagrangian ${\cal L}_Y^{\prime}$,
Eq.~(\ref{Lag-tree}), each of the couplings in the symmetry-breaking
Lagrangian ${\cal L}_Y^{\prime}$, Eq.~(\ref{BreakingLag-tree}), will
be small compared to those in ${\cal L}_Y$.

\subsection{Neutrinoless Double $\beta$ Decay}

The structure of our neutrino mass matrix leads to predictions for the
parameters measured in both neutrinoless double $\beta$-decay experiments
and $\beta$-decay experiments. From Eq.~(\ref{deltam23}), we have $m_3\simeq
\sqrt{\Delta m^2_{23}} \simeq 0.05\, \mbox{eV}$. Eqs.~(\ref{m1}) and
(\ref{m2}) then give $m_2 \simeq -2m_1 \approx b_2 \,m_3 \simeq 0.01 \,
\mbox{eV}$.

For neutrinoless double $\beta$-decay, the sum in Eq.~(\ref{nuless})
is dominated by the $i = 1,2$ terms, with $|U_{ei}^2| = {\cal O}(1)$
in each case. We estimate the experimental parameter $|m_{ee}|$,
Eq.~(\ref{nuless}), roughly, since we have neglected phases.
Assuming no cancellation, we have \beqs \label{betabetapred}
|m_{ee}|= {\cal O}(|m_2|)={\cal O}(\sqrt{\Delta m_{12}^2}) \simeq
10^{-2} \, \mbox{eV}. \eeqs This is well below the current
experimental upper bound $0.9\, \mbox{eV}$, Eq.~(\ref{nulesslimit}),
and unlikely to be accessible in next-generation experiments. This
result is not unique to our model. It would emerge in any model
leading to the normal neutrino mass hierarchy with the lightest mass
eigenvalue $m_1^2 \leq {\cal O}(\Delta m_{12}^2)$, along with the
approximate measured values of the PMNS mixing angles. The discovery
of neutrinoless double $\beta$-decay, at a higher rate, would rule
out this class of models.

For  $\beta$-decay, a similar argument gives $m_\beta \,\simeq \,
0.01 \, \mbox{eV}$. This, too, is well below the current bound,
$2.3\, \mbox{eV}$, Eq.~(\ref{betadecay}), and even less likely to be
accessible in future experiments.

\subsection{The Small Leptonic (PMNS) Mixing Angle}

The imposition of only the normal neutrino mass hierarchy within our
model has led to the ${\cal O}(b^2)$ estimate Eq.~(\ref{13}) for
$\sin^2\theta_{13}^l$. Using the numerical values of $b_1$ and $b_2$
(Table~\ref{tab:input}), we find \beq
\label{13est}\sin^2\theta_{13}^l \, \simeq \, 0.02 \,, \eeq a value
somewhat below the current experimental bound,
Eq.~(\ref{neutrinodata}), but within the anticipated range of
planned accelerator and reactor experiments. This is a rough
estimate, valid to within an ${\cal O}(1)$  multiplicative factor,
but we stress that there is no reason within our model to expect
$\sin^2\theta_{13}^l$ to be parametrically smaller than this.

The large PMNS angles have been set at tree level to their tri-bi
values $\sin^2{2\theta_{12}^l} = 8/9$ and $\sin^2{2\theta_{23}^l} =
1$ . The radiative corrections then produce shifts, of ${\cal O}(b)$
in the case of $\sin^2{2\theta_{12}^l}$, and of ${\cal O}(b^4)$ in
the case of $\sin^2{2\theta_{23}^l}$, keeping them within the
experimental range.

\section{Summary and Conclusions}

Within the general framework of an $SU(3)$ family gauge symmetry we have
described the hierarchical structure of the Yukawa couplings of the
standard-model fermions as being generated by the interplay of spontaneous
and explicit breaking of a set of global family symmetries of the standard
model, as well as the spontaneous breaking of the $SU(3)$ gauge symmetry.

Below a high scale $F$ associated with the spontaneous breaking, the
model is described by an effective field theory (EFT) containing a
small number of tree-level couplings. These determine the overall
mass scales for the fermions, breaking the separate global $U(1)$
symmetries associated with each (chiral) fermion species. These
couplings preserve a $G=SU(3)^2\times Z_3$ family symmetry, large
enough to ensure the vanishing of all the mixing angles and the
ratios between the masses of fermions with the same SM quantum
numbers.

Operators that explicitly break $G$ are present, but with  small
coefficients. The gauging of an $SU(3)\subset G$ plays the leading
role in the explicit breaking. It is the only coupling that  gives
significant loop corrections. All the mixing angles and mass
hierarchies in the quark and charged lepton sectors are computable
in terms of these corrections, restricted here to the one-loop
level.

Neutrinos play a special role in the construction. They are assumed
to couple directly (at  tree-level) with a "hidden" sector of the
theory, which otherwise communicates only through the family gauge
interaction. The generation of flavor structure (mixing matrices and
mass ratios) is controlled ultimately by the hidden sector through
the gauge interactions. Requiring that the neutrino mass matrix is
compatible with present experimental data largely determines the
internal structure of this hidden sector and renders the model
predictive.

This effective field theory approach to flavor physics has intrinsic
limitations, and cannot be used to predict (or explain) some
physical parameters such as the top mass, bottom mass, and the large
leptonic mixing angles, which are controlled by physics at energies
above the cutoff $M_F \equiv 4\pi F$. But it constitutes a
systematic tool for understanding most of the other parameters
entering the physics of flavor changing processes in the standard
model.

\vspace{1.0cm}
\begin{acknowledgments}
This work was partially supported by Department of Energy grants
DE-FG02-92ER-40704 (T.A. and Y.B.) and DE-FG02-96ER-40956 (M.P.). M.P. thanks
Michele Frigerio and Cecilia Lunardini for helpful discussions.
\end{acknowledgments}



\begin{thebibliography}{99}

\bibitem{APY2}
T.~Appelquist, Y.~Bai and M.~Piai,
  Phys.\ Lett.\ B {\bf 637}, 245 (2006)
  [arXiv:hep-ph/0603104].



\bibitem{white}
 R.~N.~Mohapatra {\it et al.},
  arXiv:hep-ph/0510213.

\bibitem{PDG}
S.~Eidelman {\it et al.}  [Particle Data Group],
         Phys.\ Lett.\ B {\bf 592}, 1 (2004).

\bibitem{SU(3)refs}

Some references on $SU(3)$ family symmetry:

M.~Bowick and P.~ Ramond, Phys. Lett.\ B {\bf 103}, 338 (1981);
D.~R.~T.~Jones, G.~L.~Kane and J.~P.~Leveille,
  Nucl.\ Phys.\ B {\bf 198}, 45 (1982);
Z.~G.~Berezhiani,
         Phys.\ Lett.\ B {\bf 150}, 177 (1985);
 Z.~G.~Berezhiani and M.~Y.~Khlopov,
  Sov.\ J.\ Nucl.\ Phys.\  {\bf 51}, 739 (1990)
  [Yad.\ Fiz.\  {\bf 51}, 1157 (1990)];
Z.~Berezhiani and A.~Rossi,
       Nucl.\ Phys.\ B {\bf 594}, 113 (2001)
       [arXiv:hep-ph/0003084];
S.~F.~King and G.~G.~Ross,
       Phys.\ Lett.\ B {\bf 520}, 243 (2001)
       [arXiv:hep-ph/0108112];
       Phys.\ Lett.\ B {\bf 574}, 239 (2003)
       [arXiv:hep-ph/0307190]
  I.~de Medeiros Varzielas and G.~G.~Ross,
  Nucl.\ Phys.\ B {\bf 733}, 31 (2006)
  [arXiv:hep-ph/0507176].
  I.~de Medeiros Varzielas, S.~F.~King and G.~G.~Ross,
  arXiv:hep-ph/0512313.




\bibitem{WZW}
      J.~Wess and B.~Zumino,
       Phys.\ Lett.\ B {\bf 37}, 95 (1971);
      E.~Witten,
       Nucl.\ Phys.\ B {\bf 223}, 422 (1983);



\bibitem{tri-bi}
  P.~F.~Harrison, D.~H.~Perkins and W.~G.~Scott,
  Phys.\ Lett.\ B {\bf 530}, 167 (2002)
  [arXiv:hep-ph/0202074].





\end{thebibliography}
\end{document}